# Electronic structure and magnetic properties of magnetically dead layers in epitaxial CoFe$_2$O$_4$/Al$_2$O$_3$/Si(111) films studied by X-ray magnetic circular dichroism

Yuki K. Wakabayashi,[1,*] Yosuke Nonaka,[2] Yukiharu Takeda,[3] Shoya Sakamoto,[2] Keisuke Ikeda,[2] Zhendong Chi,[2] Goro Shibata,[2] Arata Tanaka,[4] Yuji Saitoh,[3] Hiroshi Yamagami,[3,5] Masaaki Tanaka,[1,6] Atsushi Fujimori,[2] and Ryosho Nakane[1,7]

[1]*Department of Electrical Engineering and Information Systems,*
*The University of Tokyo, 7-3-1 Hongo, Bunkyo-ku, Tokyo 113-8656, Japan*
[2]*Department of Physics, The University of Tokyo, Bunkyo-ku, Tokyo 113-0033, Japan*
[3]*Materials Sciences Research Center, Japan Energy Atomic Agency, Sayo, Hyogo 679-5148, Japan*
[4]*Department of Quantum Matters, ADSM, Hiroshima University, Higashi-Hiroshima 739-8530, Japan*
[5]*Department of Physics, Kyoto Sangyo University, Motoyama, Kamigamo, Kita-Ku, Kyoto 603-8555, Japan*
[6]*Center for Spintronics Research Network, Graduate School of Engineering,*
*The University of Tokyo, 7-3-1 Hongo, Bunkyo-ku, Tokyo 113-8656, Japan*
[7]*Institute for Innovation in International Engineering Education,*
*The University of Tokyo, 7-3-1 Hongo, Bunkyo-ku, Tokyo 113-8656, Japan.*

[*]Current affiliation: *NTT Basic Research Laboratories, NTT Corporation, 3-1 Morinosato-Wakamiya, Atsugi, Kanagawa 243-0198, Japan*

Abstract

Epitaxial CoFe$_2$O$_4$/Al$_2$O$_3$ bilayers are expected to be highly efficient spin injectors into Si owing to the spin filter effect of CoFe$_2$O$_4$. To exploit the full potential of this system, understanding the microscopic origin of magnetically dead layers at the CoFe$_2$O$_4$/Al$_2$O$_3$ interface is necessary. In this paper, we study the crystallographic and electronic structures and the magnetic properties of CoFe$_2$O$_4$(111) layers with various thicknesses (thickness $d$ = 1.4, 2.3, 4, and 11 nm) in the epitaxial CoFe$_2$O$_4$(111)/Al$_2$O$_3$(111)/Si(111) structures using soft X-ray absorption spectroscopy (XAS) and X-ray magnetic circular dichroism (XMCD) combined with cluster-model calculation. The magnetization of CoFe$_2$O$_4$ measured by XMCD gradually decreases with decreasing thickness $d$ and finally a magnetically dead layer is clearly detected at $d$ = 1.4 nm. The magnetically dead layer has frustration of magnetic interactions which is revealed from comparison between the magnetizations at 300 and 6 K. From analysis using configuration-interaction cluster-model calculation, the decrease of $d$ leads to a decrease in the inverse-to-normal spinel structure ratio and also a decrease in the average valence of Fe at the octahedral sites. These results strongly indicate that the magnetically dead layer at the $CoFe_2O_4/Al_2O_3$ interface originates from various complex networks of superexchange interactions through the change in the crystallographic and electronic structures. Furthermore, from comparison of the magnetic properties between $d = 1.4$ and 2.3 nm, it is found that ferrimagnetic order of the magnetically dead layer at $d = 1.4$ nm is restored by the additional growth of the 0.9-nm-thick $CoFe_2O_4$ layer on it.

PACS numbers: 75.70.-i, 75.47.Lx, 75.25.-j, 73.20.-r

# I. INTRODUCTION

For the realization of Si-based spintronic devices, such as spin metal-oxide-semiconductor field-effect transistors (spin MOSFETs) [1-3], highly efficient spin injection into Si at room temperature is necessary. In a recent report [4], an epitaxial $CoFe_2O_4$(111)/$Al_2O_3$(111) bilayer was grown on a Si (111) substrate. This structure has a potential ability of injecting a highly spin-polarized current into Si, since the ideal $CoFe_2O_4$ ferrimagnetic tunnel barrier is expected to exhibit the spin filter effect with 100% efficiency at room temperature, which is caused by the spin-dependent tunnel probability due to the oppositely spin-polarized lower and higher conduction bands of $CoFe_2O_4$ with inverse spinel structure, and $CoFe_2O_4$ has a Curie temperature $T_C$ of 793 K which is high enough compared to room temperature [5-7]. Nevertheless, the experimental spin-polarization values estimated from tunnel magnetoresistance for $CoFe_2O_4$ tunnel barriers have been much less than the expectations [8,9], and many researchers point out the degradation of the electronic structure and magnetic properties of the $CoFe_2O_4$ films due to spontaneously formed structural disorder [6,10-14]. One recognized problem is that structural and chemical defects leads to the formation of mid-gap impurity states whose polarity is opposite to that of the lower conduction band of the inverse spinel structure [6,7,13]. Another problem is that anti-phase boundaries (APBs) significantly reduce the magnetization of $CoFe_2O_4$, which becomes more pronounced with decreasing thickness [10-12]. However, these problems have not been studied on $CoFe_2O_4$ films which are thin enough for electron tunneling, *i.e.*, a few nm thickness, probably because there is no established method that can clarify the structural disorder and electronic structure of such thin $CoFe_2O_4$ films. In general, it is plausible that magnetically dead layers formed at heterointerfaces degrade ferrimagnetic ordering of $CoFe_2O_4$ films with decreasing thickness, however, the magnetic properties of $CoFe_2O_4$ layers with various thicknesses have not been systematically investigated so far. Thus, it is vitally important to study the electronic structure and magnetic properties of the magnetically dead layers formed near the $CoFe_2O_4$/$Al_2O_3$ interface in epitaxial $CoFe_2O_4$(111)/$Al_2O_3$(111)/Si(111) structures to achieve highly efficient spin injection into Si.

Soft X-ray absorption spectroscopy (XAS) and X-ray magnetic circular dichroism (XMCD) are extremely sensitive tools to the local electronic structure and magnetic properties of each element in magnetic thin films [15-19], and allow us to determine the crystallographic sites and valences of cations, which determine the physical properties of oxides [20-22]. In addition, because XMCD is free from the diamagnetic signal from the substrate, one can perform accurate measurements on ultra-thin magnetically dead layers. Therefore, XMCD measurements are useful for the investigation of magnetically dead layers in oxide magnetic multilayers.

In this paper, we present the electronic structure and magnetic properties of $CoFe_2O_4$(111) layers with various thicknesses ($d$) in epitaxial $CoFe_2O_4$(111)/$Al_2O_3$(111)/Si(111) structures deduced using XAS and XMCD. The crystallographic sites and valences of the cations in the $CoFe_2O_4$ layers are determined using the experimental XMCD spectra and theoretical calculations based on the configuration-interaction (CI) cluster model [23]. We find that the magnetization gradually decreases with decreasing $d$ from 11 to 4 nm and drastically decreases with decreasing $d$ from 4 to 1.4 nm, and that the magnetization reduction is correlated to the change in the crystallographic and electronic structures. We also find that there is frustration of magnetic interactions in the magnetically dead layer. From these results, the magnetically dead layer at the $CoFe_2O_4$/$Al_2O_3$ interface probably originates from various complex networks of superexchange interactions through the change in the crystallographic and electronic structures. Furthermore, we also find that ferrimagnetic order of the magnetically dead layer at $d$ = 1.4 nm is restored by the additional growth of the 0.9-nm-thick $CoFe_2O_4$ layer on it.

## II. SAMPLE FABRICATION AND CRYSTALLOGRAPHIC ANALYSES

Epitaxial $CoFe_2O_4$(111) thin films with thickness $d$ = 1.4, 2.3, 4, and 11 nm were grown on a 2.4-nm-thick γ-$Al_2O_3$(111) buffer layer / $n^+$-Si(111) substrate using pulsed laser deposition (PLD). Figure 1 shows the schematic pictures of the sample structure [(a)] and the spinel structure [(b)] with the octahedral ($O_h$) and tetrahedral ($T_d$) sites. The small red, small blue, and large gray spheres in (b) represent the $O_h$ sites, $T_d$ sites, and oxygen anions, respectively. The blue and red arrows in (b) represent the antiferromagnetic coupling between the magnetic moment of the cations at the $T_d$ and $O_h$ sites. In the ideal inverse spinel structure, the $T_d$ sites are only occupied by the Fe cations. The regularity of the cationic distribution is represented by the inversion parameter $y$, which is a inverse-to-normal spinel structure ratio defined by the chemical formula $[Co_{1-y}Fe_y]_{Td}[Fe_{2-y}Co_y]_{Oh}O_4$. Our detailed fabrication technique and characterizations are as follows.

Phosphorus-doped Si(111) wafers with the resistivity of 2 mΩcm were used as substrates. For the growth of epitaxial γ-$Al_2O_3$ buffer layers just on the Si surfaces, we used solid-phase reaction of Al and $SiO_2$. The procedure is basically the same as that used in Ref. [24]. A Nd:YAG laser used for the PLD method was operated under the following conditions: wavelength 266 nm (fourfold wave), pulse duration 10 nsec, and pulse repetition rate 10 Hz.

First, a Si substrate was chemically cleaned with the Radio Corporation of America (RCA) method, followed by HF dip. Then, the hydrogen-terminated surface of the Si substrate was oxidized at 80°C for 5 min in a HCl : $H_2O_2$ : $H_2O$ = 3 : 1 : 1 solution to form a thin $SiO_2$ layer. Subsequently, the substrate was installed into an ultra-high vacuum deposition system having a molecular beam epitaxy (MBE) chamber and a PLD chamber. After the thermal cleaning at substrate temperature ($T_{SUB}$) = 200°C for 30 min, $T_{SUB}$ was lowered to room temperature, and then a 0.6-nm-thick Al layer was deposited on the thin $SiO_2$ surface using a Knudsen-cell in the MBE chamber. After that, the substrate was transferred to the PLD chamber via a vacuum transport chamber, and a γ-$Al_2O_3$ buffer layer was formed by solid-phase reaction of Al and $SiO_2$ at $T_{SUB}$ = 820°C for 30 min with the base pressure below $5.0\times10^{-7}$ Pa. As seen in Figs. 2(a) and (b), reflective high-energy electron diffraction (RHEED) patterns showed the Si substrate streaks together with 6-fold streaks indicating epitaxial γ-$Al_2O_3$, which were completely the same patterns as those in Ref. [24]. To obtain a more ordered surface, a 1-nm-thick epitaxial γ-$Al_2O_3$ layer was subsequently grown using a single-crystalline $Al_2O_3$ target with a rate of 0.2 nm/min. under a $O_2$ pressure ($P_{O2}$) of $1\times10^{-4}$ Pa. After the growth, the Si substrate streaks vanished, and the 6-fold γ-$Al_2O_3$ streaks became more intense and showed a 1×2 reconstruction pattern, as shown in Figs. 2(c) and (d). Then, $T_{SUB}$ was lowered to 500°C with a rate of 30°C/min under $P_{O2}$ = $1\times10^{-4}$ Pa, and finally an epitaxial $CoFe_2O_4$ layer was grown using a sintered $CoFe_2O_4$ target with a deposition rate of 0.12 nm/min. To obtain a high crystallinity, $P_{O2}$ = 10 Pa was used for this growth. Figures 2(e) and (f) show RHEED patterns of a 11-nm-thick $CoFe_2O_4$ layer, in which very sharp streaks with a 2×2 reconstruction and higher-order Laue patterns can be clearly seen. This indicates a high quality 2-dimensional epitaxial growth mode of the $CoFe_2O_4$ layer. The 6-fold symmetry with the 2×2 reconstruction was also confirmed by a low-energy electron diffraction (LEED) pattern (see Fig. S1 in Sec. I of the Supplemental Material [25]). We observed almost the same RHEED and LEED patterns for epitaxial $CoFe_2O_4$ layers with various thicknesses ($d$ = 1.4, 2.3, and 4 nm). From the analysis of the RHEED patterns, the in-plane lattice constant of the γ-$Al_2O_3$ layer was smaller by ~2 % than that of bulk γ-$Al_2O_3$, which is probably due to the lattice mismatch ~2% with respect to Si, whereas the in-plane lattice constant of the $CoFe_2O_4$ layer was smaller by ~1 % than that of bulk $CoFe_2O_4$. Thus, both layers were compressively strained in the film plane.

We evaluated the surface roughness by atomic force microscopy (AFM) and found that the surfaces of the grown γ-$Al_2O_3$ layer and the $CoFe_2O_4$ layers are continuous and very smooth, as shown in Figs. 3(a) and (b): Root mean square (RMS) values for the

γ-Al$_2$O$_3$ layer and CoFe$_2$O$_4$ layers were typically 0.25 and 0.28 nm, respectively. In Fig. 3(b), one can see facets of islands which reflect the crystallographic directions of the CoFe$_2$O$_4$ layer. Note that the RMS values of the epitaxial CoFe$_2$O$_4$ layers with various thicknesses ($d$ = 1.4, 2.3, 4, and 11 nm) were almost the same, *i. e.*, there is no thickness dependence.

Next, we characterized the crystallographic properties by X-ray diffraction (XRD). For the estimation of the out-of-plane lattice constant, the $\theta$-$2\theta$ method was used. The out-of-plane lattice constant of the γ-Al$_2$O$_3$ layer was larger by 6% than that of the bulk and, therefore, the γ-Al$_2$O$_3$ layer was tensile strained. On the other hand, the out-of-plane lattice constant of the CoFe$_2$O$_4$ layer was the same as that of the bulk. As expected from the RHEED and LEED patterns, the 6-fold symmetry of γ-Al$_2$O$_3$ and CoFe$_2$O$_4$ layers were observed by the $\phi$-scan XRD patterns for γ-Al$_2$O$_3$(311) and CoFe$_2$O$_4$(311) (see Fig. S2 in Sec. I of the Supplemental Material [25]), which indicates that one domain completely aligns to the Si substrate with the epitaxial relationship of γ-Al$_2$O$_3$[11$\bar{2}$](111) // CoFe$_2$O$_4$[11$\bar{2}$](111) // Si[11$\bar{2}$](111), whereas another domain is rotated by 60° in the (111) plane. This double domain structure is basically the same as that in Ref. [4].

Figures 4(a) and (b) show the cross-sectional high-resolution transmission electron microscopy (HRTEM) images of the CoFe$_2$O$_4$ film with $d$ = 11 nm projected along the Si <11$\bar{2}$> axis. Almost the entire region of the CoFe$_2$O$_4$ layer has an epitaxially-grown single-crystalline structure with a smooth and flat surface and interface with the γ-Al$_2$O$_3$ buffer layer, as expected from the RHEED patterns, the AFM images, and the XRD analysis. One can observe a ~2.1-nm-thick SiO$_x$ interfacial layer, which might be caused by the high O$_2$ pressure of 10 Pa during the growth of the CoFe$_2$O$_4$ layer. The orange dashed lines represent APBs, which are growth defects of the cation sublattice in the spinel structure [10-12]: The oxygen lattice remains unchanged across an APB whereas the cation sublattice is shifted by the <220> translation vector [12]. In the previous reports on CoFe$_2$O$_4$/MgO [26] and CoFe$_2$O$_4$/α-Al$_2$O$_3$ [12], the density of APBs decreases with increasing thickness of the CoFe$_2$O$_4$ layer since APBs, which are mainly introduced at the heterointerface, vanish monotonically with increasing thickness. Contrary to these, in our case, the heterointerface CoFe$_2$O$_4$/γ-Al$_2$O$_3$ is not clearly identified in Fig. 4(b) and thus it is not the main source of APBs. As seen in Fig. 4(b), two APBs penetrate the whole CoFe$_2$O$_4$ layers and the other two APBs vanish in the γ-Al$_2$O$_3$ buffer layer. Since the APB distribution is basically the same in a wide area, it seems that the density of APBs does not depend on the thickness of the CoFe$_2$O$_4$ layer.

**III. XAS AND XMCD MEASUREMENTS**

We performed XAS and XMCD measurements at the soft X-ray beamline BL23SU of SPring-8 with a twin-helical undulator of in-vacuum type [27], which allows us to perform efficient and accurate measurements of XMCD with various incident photon energies and magnetic fields. The monochromator resolution was $E/\Delta E > 10000$. XMCD spectra were obtained by reversing the photon helicity at each energy point and were recorded in the total-electron-yield (TEY) mode. To eliminate experimental artifacts, we averaged XMCD spectra taken for both positive and negative magnetic fields applied perpendicular to the film surface. The direction of the incident X-rays was also perpendicular to the film surface. Backgrounds of the XAS spectra at the Fe and Co $L_{2,3}$-edges were subtracted from the raw data, assuming that they are hyperbolic tangent functions.

Figure 5 shows the Fe and Co $L_{2,3}$-edge XAS [$\mu^-$, $\mu^+$, and $(\mu^+ + \mu^-)/2$] and XMCD ($\mu^+ - \mu^-$) spectra for the CoFe$_2$O$_4$ film with $d = 11$ nm measured at 6 and 300 K with magnetic fields $\mu_0H = 0$, 1, and 7 T. The spectra with both $\mu_0H = 0$ and 1 T were measured after the application of $\mu_0H = 7$ T. Here, $\mu^+$ and $\mu^-$ denote the absorption coefficients for the photon helicities parallel and antiparallel to the Co 3$d$ majority spin direction, respectively. These spectra show the multiplet structures which are a characteristic of the localized 3$d$ state of Fe and Co cations in oxides [20-22]. For both the Fe and Co $L_{2,3}$-edges, the XMCD spectra with $\mu_0H = 7$ T are identical between 6 and 300 K, indicating that the $T_\mathrm{C}$ of the CoFe$_2$O$_4$ film is sufficiently higher than 300 K. The XMCD intensities with $\mu_0H = 0$ T originate from the remanent-spontaneous magnetization. The insets of Figs. 5 (c) and (d) show the magnified plots of the XMCD spectra at the $L_3$-edges normalized at 709.2 and 778.2 eV, respectively. For both the Fe and Co $L_3$-edges, the normalized XMCD spectra for various $H$ are identical with each other, indicating that the XMCD signals come from a single magnetic phase. Since the same features were also observed for the CoFe$_2$O$_4$ films with $d = 1.4$, 2.3, and 4 nm (see Section II of the Supplemental Material [25]), all the CoFe$_2$O$_4$ films with various thicknesses are magnetically uniform. The Fe $L_3$-edge XMCD spectrum has two negative peaks at 707.38 (peak $a$) and 709.18 eV (peak $c$) and a positive peak at 708.48 eV (peak $b$). It is well known that the XMCD peaks at $a$, $b$, and $c$ mainly come from Fe cations at Fe$^{2+}$ ($O_h$), Fe$^{3+}$ ($T_d$), and Fe$^{3+}$ ($O_h$) sites, respectively [20,28], where $O_h$, $T_d$, and the superscript number denote octahedral site, tetrahedral site, and the valence of the Fe cations, respectively [see Fig. 1(b)]. Our calculation also supports these assignments, as we shall show below [Figs. 7(a) and (b)]. The comparable peak height of peak $b$ and peak $c$ indicates that the amount of the Fe$^{3+}$ ($T_d$) cations is comparable to that of the Fe$^{3+}$ ($O_h$) cations. This result means that the CoFe$_2$O$_4$ layer has a high inversion parameter $y$. On the other hand, the XMCD signals at the Co $L_3$-edge are mostly negative. Thus, it was found that the magnetic moments of the $Fe^{3+}$ ($T_d$) cations and Co cations have an antiparallel configuration, as shown in Fig. 1(b), which is a characteristic feature of the Co ($O_h$) cations in inverse spinel ferrites [20].

Figure 6(a) shows the Fe $L_{2,3}$-edge XMCD spectra for the $CoFe_2O_4$ films with $d =$ 1.4, 2.3, 4, and 11 nm, which were measured at 300 K with $\mu_0 H = 7$ T. The maximum XMCD intensity at the Fe $L_3$-edge decreases by 33% in going from 11 to 4 nm, and then it drastically decreases by 79% in going from 4 to 1.4 nm. The normalized XMCD intensity at the peak $b$, which comes from the $Fe^{3+}$ ($T_d$) cations, decreases with decreasing $d$ [Fig. 6(b)]. Thus, the drastic reduction of the magnetization is seemingly associated with the decrease in the inverse-to-normal spinel structure ratio.

To clarify correlation between the magnetization and the crystallographic and electronic structures quantitatively, we analyzed the crystallographic sites and valences of Fe and Co cations using the experimental XMCD spectra and cluster-model calculation. It has been well recognized that the XAS and XMCD spectra of transition-metal oxides strongly depend on the $3d$ electron configurations, crystal field, spin-orbit coupling, and electron-electron interaction within the transition-metal cation, and the hybridization of $3d$ electrons with other valence electrons. Taking into account these effects, XAS and XMCD spectra for Fe and Co with a specific site and valence can be calculated by employing the configuration-interaction (CI) cluster model [23]. In the cluster-model calculation, we adopted empirical relationships between the on-site Coulomb energy $U_{dd}$ and the $3d$-$2p$ hole Coulomb energy $U_{dc}$: $U_{dc} / U_{dd} = 1.25$ [28], and between the Slater-Koster parameters $pd\sigma$ and $pd\pi$: $pd\sigma / pd\pi = -2.17$ [29]. The hybridization strength between O $2p$ orbitals $T_{pp}$ was fixed to be 0.7 eV (for $O_h$ site) and 0 eV (for $T_d$ site) [23,28]. The 80% of ionic Hartree-Fock values were used for Slater integrals. In the cluster-model calculation for Fe cations, model parameters were determined as follows: We used the reported $Fe_3O_4$ values of the $U_{dd}$ [28]. Charge-transfer energy $\Delta$, which is defined as the energy required to transfer an electron from O $2p$ to Fe $3d$ orbitals, the crystal field splitting $10Dq$, and the Slater-Koster parameters $pd\sigma$ between Fe $3d$ and O $2p$ were determined by curve fitting to the experimental Fe $L_{2,3}$-edge XAS and XMCD spectra (particularly for the $L_3$-edges) measured at 300 K with $\mu_0 H = 7$ T of the $CoFe_2O_4$ film with $d = 11$ nm. The parameter values determined for the Fe cations are listed in Table 1.

Figures 7(a) and (b) show the calculated Fe $L_{2,3}$-edge XAS and XMCD spectra for the $Fe^{3+}$ ($O_h$), $Fe^{3+}$ ($T_d$), and $Fe^{2+}$ ($O_h$) cations, respectively, using the parameters in Table 1. The spin magnetic moment $m_{spin}$ and the orbital magnetic moment $m_{orb}$ are also calculated within the CI cluster model based on the above parameters for the $Fe^{3+}$ ($O_h$), $Fe^{3+}$ ($T_d$), and $Fe^{2+}$ ($O_h$) cations, and these are summarized in Table 2. Figures 7(c) and (d) show the experimental Fe $L_{2,3}$-edge XAS and XMCD spectra measured at 300 K with $\mu_0 H = 7$ T for the $CoFe_2O_4$ films with $d$ = 1.4, 2.3, 4, and 11 nm, and the corresponding curve fittings (particularly for the $L_3$-edges) with the weighted sum of the calculated spectra shown in Figs. 7(a) and (b). The experimental spectra are well reproduced by the weighted sum of the calculated spectra, including the characteristic kink structure at around 708 eV in the experimental XMCD spectra. These results give strong evidence that the Fe cations only occupy the $Fe^{3+}$ ($O_h$), $Fe^{3+}$ ($T_d$), and $Fe^{2+}$ ($O_h$) sites. From these fits, we can obtain the magnetic moment of the $Fe^{3+}$ ($O_h$), $Fe^{3+}$ ($T_d$), and $Fe^{2+}$ ($O_h$) cations at 300 K with $\mu_0 H = 7$ T and the site occupancies for the Fe cations.

Figure 8 shows the $d$ dependence of the site occupancies for $Fe^{3+}$ ($O_h$), $Fe^{3+}$ ($T_d$), and $Fe^{2+}$ ($O_h$) estimated from the same fitting procedure. In the figure, the inversion parameter $y$ is also shown. The $y$ value gradually decreases with decreasing $d$ and finally drops to 0.54 at $d$ = 1.4 nm. Another striking feature is that the site occupancy for $Fe^{2+}$ ($O_h$) at $d$ = 1.4 nm is twice as large as those in the thicker samples, whereas the site occupancy for $Fe^{3+}$ ($O_h$) is almost independent of $d$. Thus, the decrease of $d$ leads to the decrease in the inverse-to-normal spinel structure ratio and also decrease in the average valence of the Fe cations at the $O_h$ sites.

In the calculation for the Co cations using the CI cluster model, the parameters are determined as follows: The $\Delta$, $10Dq$, $U_{dd}$, and $pd\sigma$ were determined by the curve fitting for the experimental Co $L_{2,3}$-edge XAS and XMCD spectra measured at 300 K with $\mu_0 H = 7$ T of the $CoFe_2O_4$ film with $d$ = 11 nm. The determined parameter values for the Co cations are also listed in Table 1.

Figures 9(a) and (b) show the calculated Co $L_{2,3}$-edge XAS and XMCD spectra for the $Co^{2+}$ ($O_h$), $Co^{2+}$ ($T_d$), and $Co^{3+}$ ($O_h$) cations, respectively, using the parameters in Table 1. The spin magnetic moment $m_{spin}$ and the orbital magnetic moment $m_{orb}$ are also calculated within the cluster model using the above parameters for the $Co^{2+}$ ($O_h$), $Co^{2+}$ ($T_d$), and $Co^{3+}$ ($O_h$) cations, and these are also summarized in Table 2. Figures 9(c) and (d) show the experimental Co $L_{2,3}$-edge XAS and XMCD spectra measured at 300 K with $\mu_0 H = 7$ T for the $CoFe_2O_4$ films with $d$ = 1.4, 2.3, 4, and 11 nm, and the weighted sum of the calculated spectra shown in Figs. 9(a) and (b). Here, the site occupancies for the Co cations are calculated from those for the Fe cations shown in Fig. 8 so that the charge neutrality is fulfilled and the number ratio of the $O_h$ sites to the $T_d$ sites is 2 in the $CoFe_2O_4$ layer. From these calculations, we obtain the magnetization of the $Co^{2+}$ ($O_h$), $Co^{2+}$ ($T_d$), and $Co^{3+}$ ($O_h$) cations at 300 K with $\mu_0 H = 7$ T. The experimental XAS and XMCD spectra with $d = 11$ nm are well reproduced by the weighted sum of the calculated spectra as shown in Figs. 9(c) and (d), which indicates that most Co cations occupy the $Co^{2+}$ ($O_h$), $Co^{2+}$ ($T_d$), and $Co^{3+}$ ($O_h$) sites in the $CoFe_2O_4$ films with $d = 11$ nm. However, the discrepancy between the experimental spectra and the weighted sum of the calculated spectra increases with decreasing $d$, and finally the experimental XMCD spectra with $d = 1.4$ nm could not be reproduced by the weighted sum of the calculated spectra [Fig. 9(d)]. This result suggests that there are other types of Co cations such as low spin Co cations [30], Co cations at the trigonal prism sites [31], and Co cations under strong local distortion and/or that the ratio of the $O_h$ site to the $T_d$ site changes from 2 [32] near the $CoFe_2O_4/Al_2O_3$ interface.

Figures 10(a)-(d) show the XMCD - $H$ curves at the Fe $L_3$-edge for the $CoFe_2O_4$ films with $d = 11, 4, 2.3$, and $1.4$ nm, respectively, in which red curves and blue curve are the results measured at 300 K and 6 K, respectively. Here, we have scaled the vertical axis so that the XMCD intensity at 300 K with $\mu_0 H = 7$ T represents the sum of the magnetizations of the Fe and Co cations. Taking into the antiferromagnetic coupling between the magnetic moment of the cations at $T_d$ and $O_h$ sites, the sum of the magnetizations was calculated from the sum of the magnetic moments of the $Fe^{3+}$ ($O_h$), $Fe^{3+}$ ($T_d$), $Fe^{2+}$ ($O_h$), $Co^{2+}$ ($O_h$), $Co^{2+}$ ($T_d$), and $Co^{3+}$ ($O_h$) cations estimated in Figs. 7(c), 7(d), 9(c), and 9(d) with the ratio of the site occupancies shown in Fig. 8. In all the samples, the normalized XMCD - $H$ curves measured at various energies at the Fe and Co $L_3$-edges are identical, which confirms again that the XMCD signals come from a single magnetic component (see Section III of the Supplemental Material [25]). One can clearly see hysteresis in the $CoFe_2O_4$ film with $d = 11$ nm, which indicates that $T_C$ is higher than 300 K. For $d \leq 4$ nm, the XMCD - $H$ curve at 300 K does not have hysteresis, its linearity increases with decreasing $d$, and finally it shows a nearly straight line for $d = 1.4$ nm. These features indicate that random magnetization configurations in zero field [11,26] are induced by various complex networks of superexchange interactions through the change in the crystallographic and electronic structures shown in Fig. 8. Since we detected signals originating from magnetic ions located within the depth of 2–3 nm from the surface [33], the XMCD spectra for $d = 1.4$ and 2.3 nm reflect the magnetic properties at the $CoFe_2O_4/Al_2O_3$ interface. Thus, the nearly linear XMCD - $H$ relation for $d = 1.4$ nm at 300 K probably come from a magnetically dead layer formed at the $CoFe_2O_4/Al_2O_3$ interface. To study this further, XMCD - $H$ for $d = 1.4$ nm was also measured at 6 K, as shown by a blue curve in Fig. 10(d). Although the curve is slightly nonlinear, the magnetization at 7 T was increased by only 2.5 times when temperature was decreased from 300 to 6 K, which means that the magnetization does not originates only from paramagnetic component and that there is the frustration of magnetic interactions induced by various complex networks of superexchange interactions in the magnetically dead layer. On the other hand, the XMCD – $H$ curve for $d = 2.3$ nm is less linear than that for $d = 1.4$ nm. Since all the XMCD signals come from the single magnetic phase as described above, ferrimagnetic order of the magnetically dead layer at $d = 1.4$ nm is restored by the additional growth of the 0.9-nm-thick $CoFe_2O_4$ layer on it. Thus, the magnetic properties of the magnetically dead layer at the $CoFe_2O_4/Al_2O_3$ interface are found to be determined not only by the crystallographic and electronic structures near the interface, but also by the thickness of the $CoFe_2O_4$ layer.

In the following, we discuss our results in comparison with previous studies by other groups. To our knowledge, it is well-recognized that APBs, which are growth defects in the cation sublattice, induces 180° Fe-O-Fe antiferromagnetic superexchange interactions and reduces the magnetization of magnetite $Fe_3O_4$ films [10,11]. Moussy *et al*. [12] reported that the density of APBs in epitaxial $Fe_3O_4(111)$ films increased with decreasing thickness, and this feature led to the decrease of the magnetization with decreasing thickness; the magnetization at 300 K with $\mu_0H = 1$ T decreased by ~15% with decreasing thickness from 15 to 5 nm. In our case, as shown in Fig. 11, the sum of the magnetizations of the Fe and Co cations in the $CoFe_2O_4$ film at 7 T decreased by 43% from $d = 11$ to 4 nm. Thus, the decrease ratio of the magnetization in the $CoFe_2O_4$ films in the large thickness range (11 to 4 nm) is slightly larger but comparable to that in the $Fe_3O_4$ films. Furthermore, the absolute value of the slope of the magnetization-$d$ curve (Fig. 11) of the $CoFe_2O_4$ films in the small thickness range of 4 to 1.4 nm (58 emu/cc·nm) is significantly larger than that in the large thickness range of 11 to 4 nm (20 emu/cc·nm). The APBs model described above may indeed be applied in the large thickness range (11 to 4 nm), but it cannot be applied in the small thickness range (4 to 1.4 nm), since the decrease ratio in the large thickness range is comparable to the value reported for the $Fe_3O_4$ films but the decrease ratio in the small thickness range is significantly larger than that in the small thickness range. In our samples, the density of the APBs does not clearly depend on the thickness of the $CoFe_2O_4$ layer from the cross-sectional TEM image in Fig. 4(b), whereas the crystallographic and electronic structures drastically change in the thickness range of 4 to 1.4 nm as shown in Fig. 8. Therefore, it is reasonable to conclude that the crystallographic and electronic structures predominantly affect the magnetic order in the $CoFe_2O_4$ region very close to the $CoFe_2O_4/Al_2O_3$ interface (within 4 nm).

## IV. CONCLUSION

We have investigated the electronic structure and magnetic properties of the $CoFe_2O_4(111)$ layers with various thicknesses ($d$ = 1.4, 2.3, 4, and 11 nm) in the epitaxial $CoFe_2O_4/Al_2O_3/Si$ structures. The XAS and XMCD spectra revealed that the magnetization gradually decreases with decreasing $d$ and finally the magnetically dead layer was clearly detected at $d$ = 1.4 nm. The magnetically dead layer has frustration of magnetic interactions which was characterized by the comparison of magnetizations measured at 300 and 6 K. Using these experimental XAS and XMCD spectra, the sites and valences of Fe cations were estimated by employing the CI cluster model. We found that the decrease of $d$ leads to the decreases in the inverse-to-normal spinel structure ratio and in the average valence of the Fe cations at the $O_h$ sites. From the experimental results and calculation, it was found that the magnetically dead layer at the $CoFe_2O_4/Al_2O_3$ interface probably originates from various complex networks of superexchange interactions through the change in the crystallographic and electronic structures. Furthermore, we have also found that ferrimagnetic order of the magnetically dead layer at $d$ = 1.4 nm is restored by the additional growth of the 0.9-nm-thick $CoFe_2O_4$ layer on it. These findings are important for the design of Si-based spintronic devices using spinel ferrites.

**ACKNOWLEDGEMENTS**

This work was partly supported by Grants-in-Aid for Scientific Research (Grants No. 26289086 and No. 15H02109), including the Project for Developing Innovation Systems from Ministry of Education, Culture, Sports, Science and Technology (MEXT), the Cooperative Research Project Program of Research Institute of Electrical Communication (RIEC), Tohoku University, and the Spintronics Research Network of Japan (SRNJ). This work was performed under the Shared Use Program of Japan Atomic Energy Agency (JAEA) Facilities (Proposal No. 2016A-E27) with the approval of the Nanotechnology Platform Project supported by MEXT. The synchrotron radiation experiments were performed at the JAEA beamline BL23SU in SPring-8 (Proposal No. 2016A3831). Y. K. W. acknowledges financial support from Japan Society for the Promotion of Science (JSPS) through the Program for Leading Graduate Schools (MERIT) and the JSPS Research Fellowship Program for Young Scientists. S. S. acknowledges financial support from JSPS through the Program for Leading Graduate Schools (ALPS).

**References**
[1] S. Sugahara and M. Tanaka, Appl. Phys. Lett. **84**, 2307 (2004).
[2] T. Tahara, H. Koike, M. Kameno, T. Sasaki, Y. Ando, K. Tanaka, S. Miwa, Y. Suzuki, and M. Shiraishi, Appl. Phys. Express **8**, 113004 (2015).
[3] R. Nakane, T. Harada, K. Sugiura, and M. Tanaka, Jpn, J. Appl. Phys. **49**, 113001 (2010).
[4] R. Bachelet, P. de Coux, B. Warot-Fonrose, V. Skumryev, G. Niu, B. Vilquin, G. Saint-Girons, and F. Sánchez, CrystEngComm **16**, 10741 (2014).
[5] Y. Suzuki, R. B. van Dover, E. M. Gyorgy, J. M. Phillips, V. Korenivski, D. J. Werder, C. H. Chen, R. J. Cava, J. J. Krajewski, and W. F. Peck, Jr., and K. B. Do, Appl. Phys. Lett. **68**, 714 (1996).
[6] A. V. Ramos, M.-J. Guitted, J.-B. Moussy, R. Mattana, C. Deranlot, F. Petroff, and C. Gatel, Appl. Phys. Lett. **91**, 122107 (2007).
[7] Z. Szotek, W. M. Temmerman, D. Kodderitzsch, A. Svane, L. Petit, and H. Winter, Phys. Rev. B **74**, 174431 (2006).
[8] J.S Moodera, T. S. Samtos, and T. Nagahama, J. Phys. Condens. Matter **19**, 165202 (2007).
[9] J-B. Moussy, J. Phys. D: Appl. Phys. **46** 143001 (2013).
[10] F. C. Voogt, T. T. M. Palstra, L. Niesen, O. C. Rogojanu, M. A. James, and T. Hibma, Phys. Rev. B **57**, R8107 (1998).
[11] D. T. Margulies, F. T. Parker, M. L. Rudee, F. E. Spada, J. N. Chapman, P. R. Aitchison, and A. E. Berkowitz, Phys. Rev. Lett **79**, 5162 (1997).
[12] J.-B. Moussy, S. Gota, A. Bataille, M.-J. Guittet, M. Gautier-Soyer, F. Delille, B. Dieny, F. Ott, T. D. Doan, P. Warin, P. Bayle-Guillemaud, C. Gatel, and E. Snoeck, Phys. Rev. B **70**, 174448 (2004).
[13] Y. K. Takahashi, S. Kasai, T. Furubayashi, S. Mitani, K. Inomata, and K. Hono, Appl. Phys. Lett. **96**, 072512 (2010).
[14] T. Niizeki, Y. Utsumi, R. Aoyama, H. Yanagihara, J. -I. Inoue, Y. Yamasaki, H. Nakao, K. Koike, and E. Kita, Appl. Phys. Lett. **103**, 162407 (2013).
[15] B. T. Thole, P. Carra, F. Sette, and G. van der Laan, Phys. Rev. Lett. 68, 1943 (1992).
[16] C. T. Chen, Y. U. Idzerda, H. -J. Lin, N. V. Smith, G. Meigs, E. Chaban, G. H. Ho, E. Pellegrin, and F. Sette, Phys. Rev. Lett. 75, 152 (1995).
[17] J. Stohr and H. Konig, Phys. Rev. Lett. 75, 3748 (1995).
[18] Y. K. Wakabayashi, S. Sakamoto, Y. Takeda, K. Ishigami, Y. Takahashi, Y. Saitoh, H. Yamagami, A. Fujimori, M. Tanaka, and S. Ohya. Sci. Rep. **6**, 23295 (2016).
[19] Y. K. Wakabayashi, R. Akiyama, Y. Takeda, M. Horio, G. Shibata, S. Sakamoto, Y. Ban, Y. Saitoh, H. Yamagami, A. Fujimori, M. Tanaka, and S. Ohya, Phys. Rev. B **95**, 014417 (2017).
[20] S. Matzen, J.-B. Moussy, R. Mattana, F. Petroff, C. Gatel, B. Warot-Fonrose, J. C. Cezar, A. Barbier, M.-A. Arrio, and Ph. Sainctavit, Appl. Phys. Lett. **99**, 052514 (2011).
[21] S. Matzen, J.-B. Moussy, R. Mattana, K. Bouzehouane, C. Deranlot, F. Petroff, J. C. Cezar, M.-A. Arrio, Ph. Sainctavit, C. Gatel, B. Warot-Fonrose, and Y. Zheng, Phys. Rev. B **83**, 184402 (2011).

[22] B. B. Nelson-Cheeseman, R. V. Chopdekar, J. M. Iwata, M. F. Toney, E. Arenholz, and Y. Suzuki, Phys. Rev. B **82**, 144419 (2010).

[23] A. Tanaka and T. Jo, J. Phys. Soc. Jpn **63**, 2788 (1994).

[24] Y. Jung, H. Miura and M. Ishida, Jpn. J. Appl. Phys. **38**, 2333 (1999).

[25] See Supplemental Material at [URL] for the LEED and XRD measurements, sample characterizations, XAS and XMCD spectra with the thickness $d$ = 1.4, 2.3 and 4 nm, and XMCD-$\mu_0 H$ curves measured at the Fe and Co $L_3$-edges.

[26] W. Eerenstein, T. T. M. Palstra, T. Hibma, and S. Celotto, Phys. Rev. B **66**, 201101(R) (2002).

[27] Y. Saitoh, Y. Fukuda, Y. Takeda, H. Yamagami, S. Takahashi, Y. Asano, T. Hara, K. Shirasawa, M. Takeuchi, T. Tanaka, and H. Kitamura, J. Synchrotron Rad. **19**, 388 (2012).

[28] J. Chen, D. J. Huang, A. Tanaka, C. F. Chang, S. C. Chung, W. B. Wu, and C. T. Chen, Phys. Rev. B **69**, 085107 (2004).

[29] Y. Fukuma, T. Taya, S. Miyawaki, T. Irisa, H. Asada, and T. Koyanagi J. Appl. Phys. **99**, 08D508 (2006).

[30] M. Oku and K. Hirokawa, J. Electron Spectrosc. Relat. Phenom. **8**, 475 (1976).

[31] S. Niitaka, H. Kageyama, M. Kato, K. Yoshimura, K. Kosuge, J. Solid State Chem. **146**, 137 (1999).

[32] C. F. Chang, Z. Hu, S. Klein, X. H. Liu, R. Sutarto, A. Tanaka, J. C. Cezar, N. B. Brookes, H.-J. Lin, H. H. Hsieh, C. T. Chen, A. D. Rata, and L. H. Tjeng, Phys. Rev. X **6**, 041011 (2016).

[33] R. Nakajima, J. Stohr, and Y. U. Idzerda, Phys. Rev. B **59**, 6421 (1999).

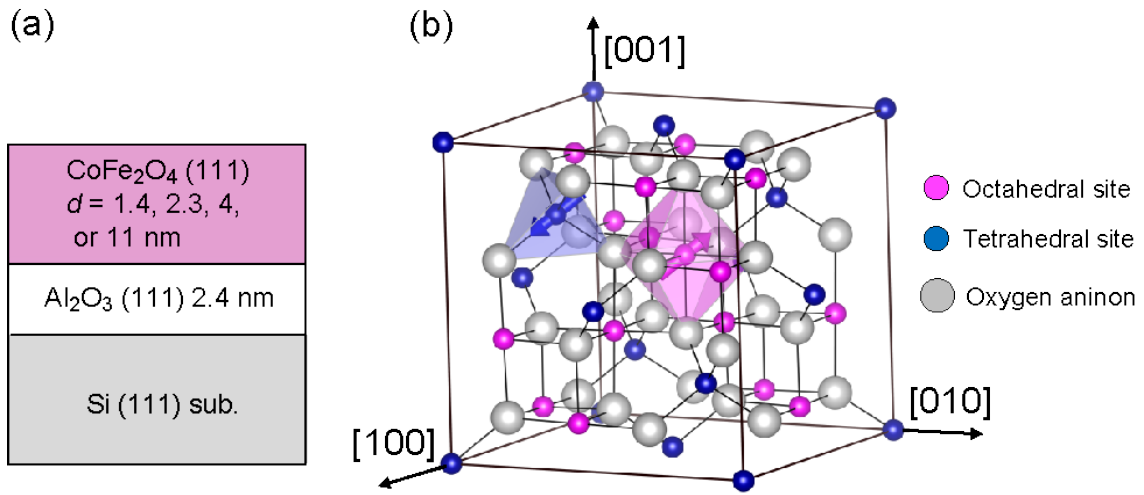

FIG. 1. (a)(b) Schematic pictures of the sample structure [(a)] and the spinel structure [(b)] with the octahedral ($O_h$) and tetrahedral ($T_d$) sites. The small red, small blue, and large gray spheres in (b) represent the $O_h$ sites, $T_d$ sites, and oxygen anions, respectively. The blue and red arrows in (b) represent the antiferromagnetic coupling between the magnetic moment of the cations at the $T_d$ and $O_h$ sites.

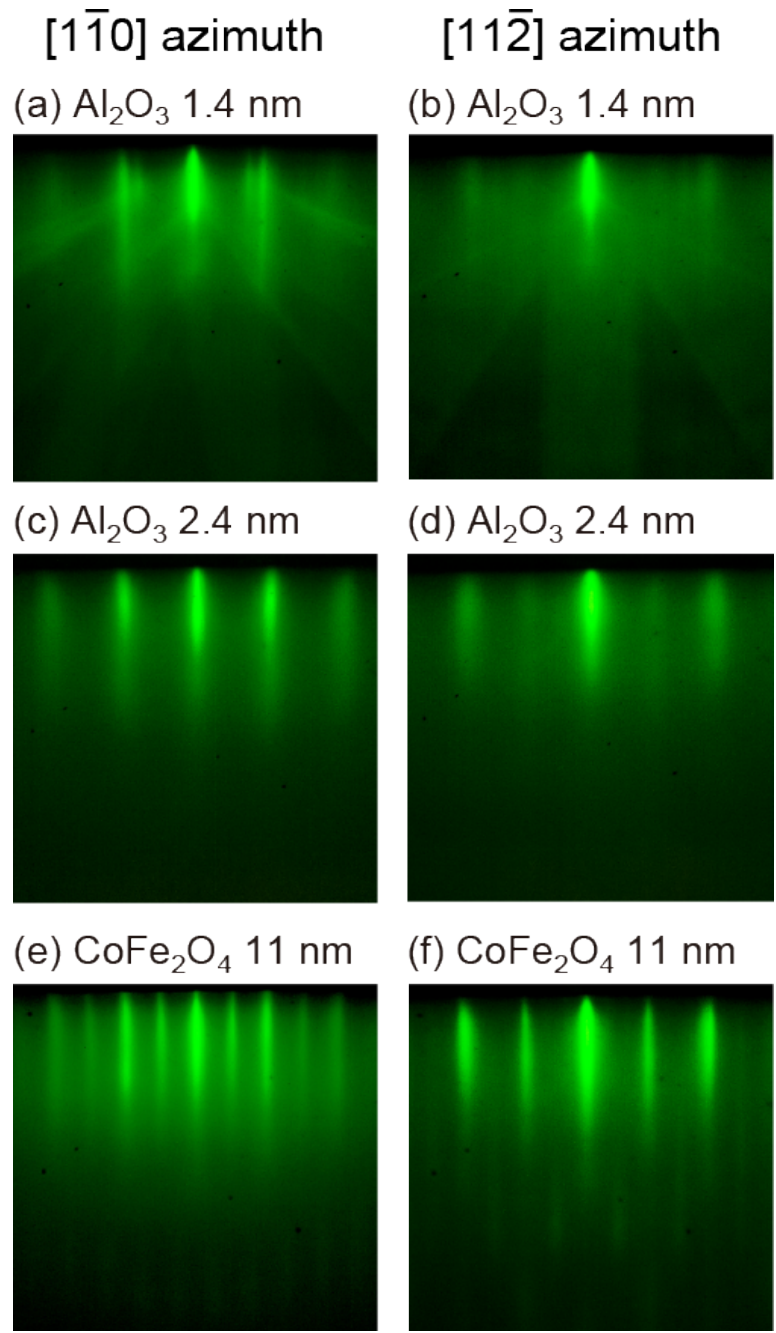

FIG. 2. Reflective high-energy electron diffraction (RHEED) patterns of an epitaxial CoFe$_2$O$_4$/γ-Al$_2$O$_3$(11$\underline{1}$)/Si(111) structure, where the electron incidences are along the [1$\bar{1}$0] [(a)(c)(e)] and [11$\bar{2}$] [(b)(d)(f)] directions of the Si substrate. (a)(b) After the growth of 1.4-nm-thick γ-Al$_2$O$_3$(111) on the Si surface by annealing. (c)(d) After the growth of an additional 1-nm-thick γ-Al$_2$O$_3$ layer by PLD. (e)(f) After the growth of a 11-nm-thick CoFe$_2$O$_4$ layer by PLD.

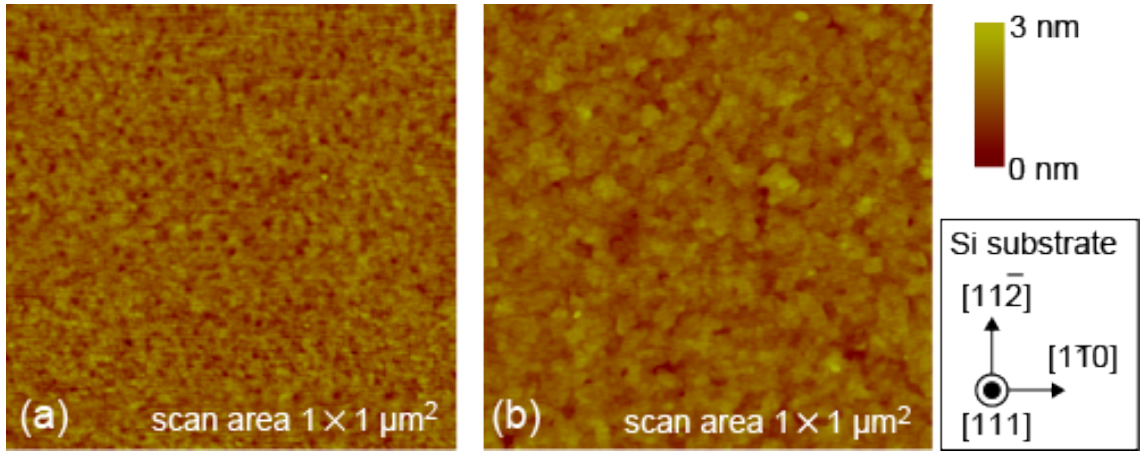

FIG. 3. Atomic force microscopy (AFM) images of the 2.4-nm-thick γ-Al$_2$O$_3$(111) layer [(a)] and the 11-nm-thick CoFe$_2$O$_4$ layer [(b)] in an epitaxial CoFe$_2$O$_4$/γ-Al$_2$O$_3$(111)/Si(111) structure. The scan area is 1×1 μm$^2$. The crystallographic directions of the Si substrate are also shown.

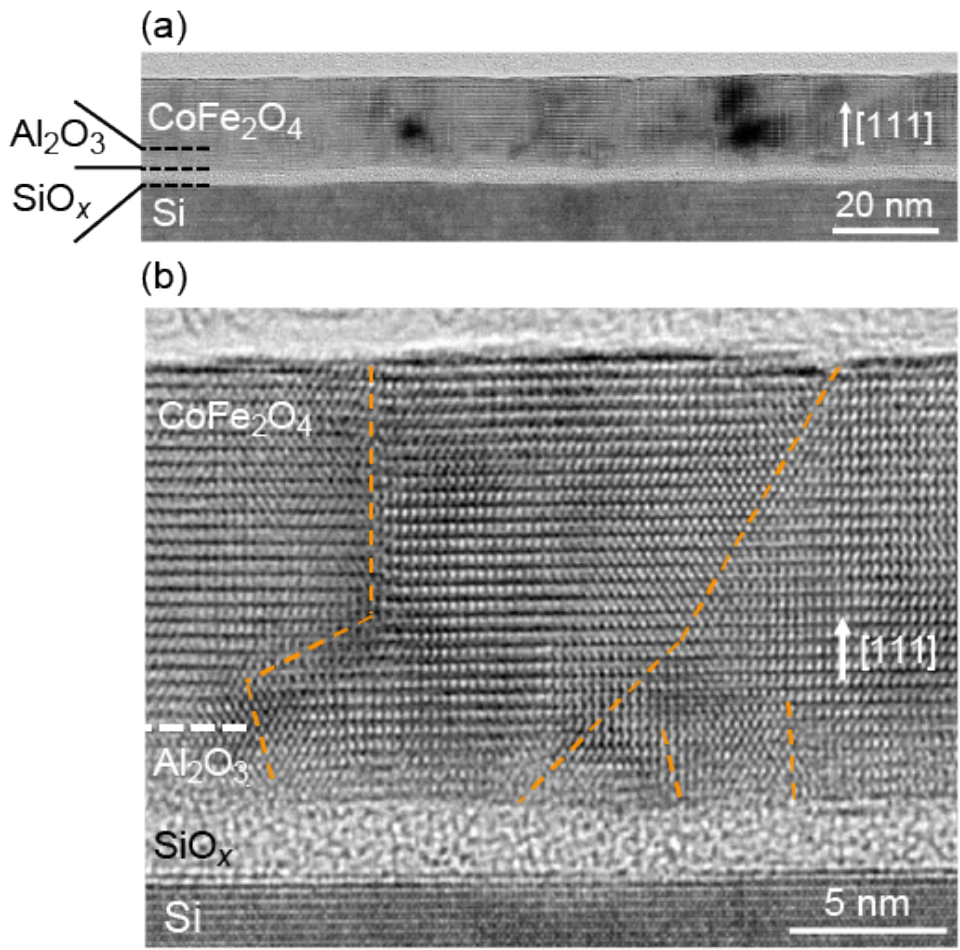

FIG. 4. HRTEM lattice image of the CoFe$_2$O$_4$ film with $d$ = 11 nm projected along the Si <11$\bar{2}$> axis. (b) Magnified image of (a). The orange dashed lines represent APBs. The oxygen lattice remains unchanged across the APB while the cation sublattice shifts by a <220> translation vector.

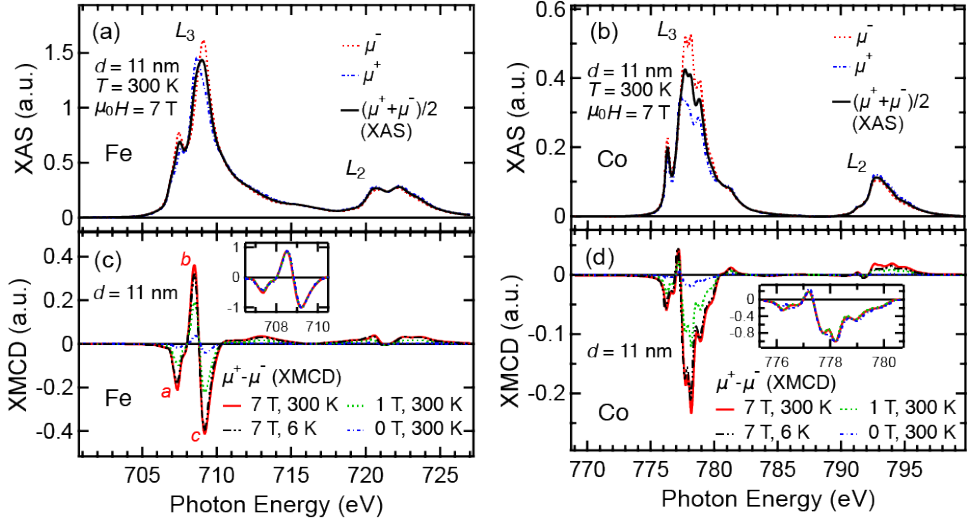

FIG. 5. (a) Fe and (b) Co $L_{2,3}$-edge XAS $[(\mu^+ + \mu^-)/2]$ spectra for the CoFe$_2$O$_4$ film with $d$ = 11 nm measured at 300 K with a magnetic field $\mu_0 H$ = 7 T applied perpendicular to the film surface. (c) Fe and (d) Co $L_{2,3}$-edge XMCD (= $\mu^+ - \mu^-$) spectra for the CoFe$_2$O$_4$ film with $d$ = 11 nm measured at 6 and 300 K with magnetic fields $\mu_0 H$ = 0, 1, and 7 T applied perpendicular to the film surface. The insets of (c) and (d) show magnified plots of the spectra at the $L_3$-edges normalized at 709.2 and 778.2 eV, respectively.

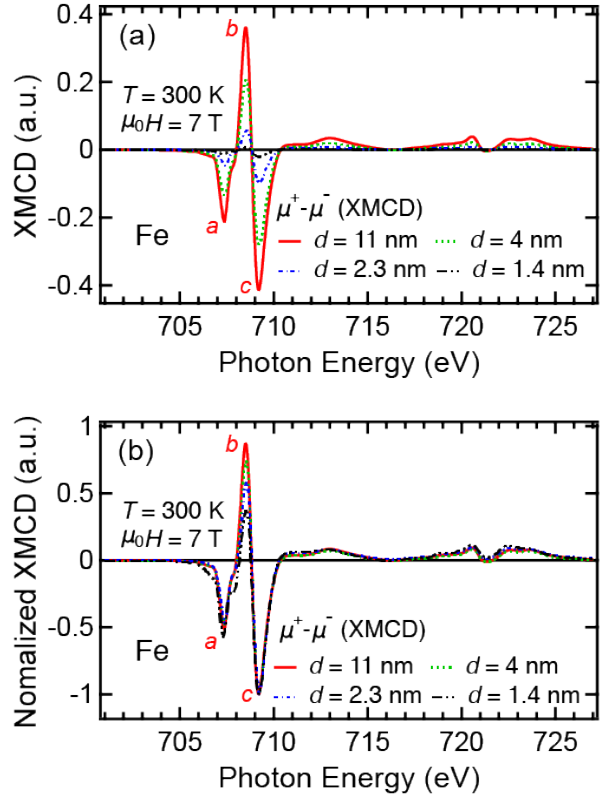

FIG. 6. (a) Fe $L_{2,3}$-edge XMCD spectra for the CoFe$_2$O$_4$ films with $d$ = 1.4, 2.3, 4, and 11 nm measured at 300 K with a magnetic field $\mu_0 H$ = 7 T applied perpendicular to the film surface. (b) Spectra in (a) normalized to the intensity at 709.2 eV.

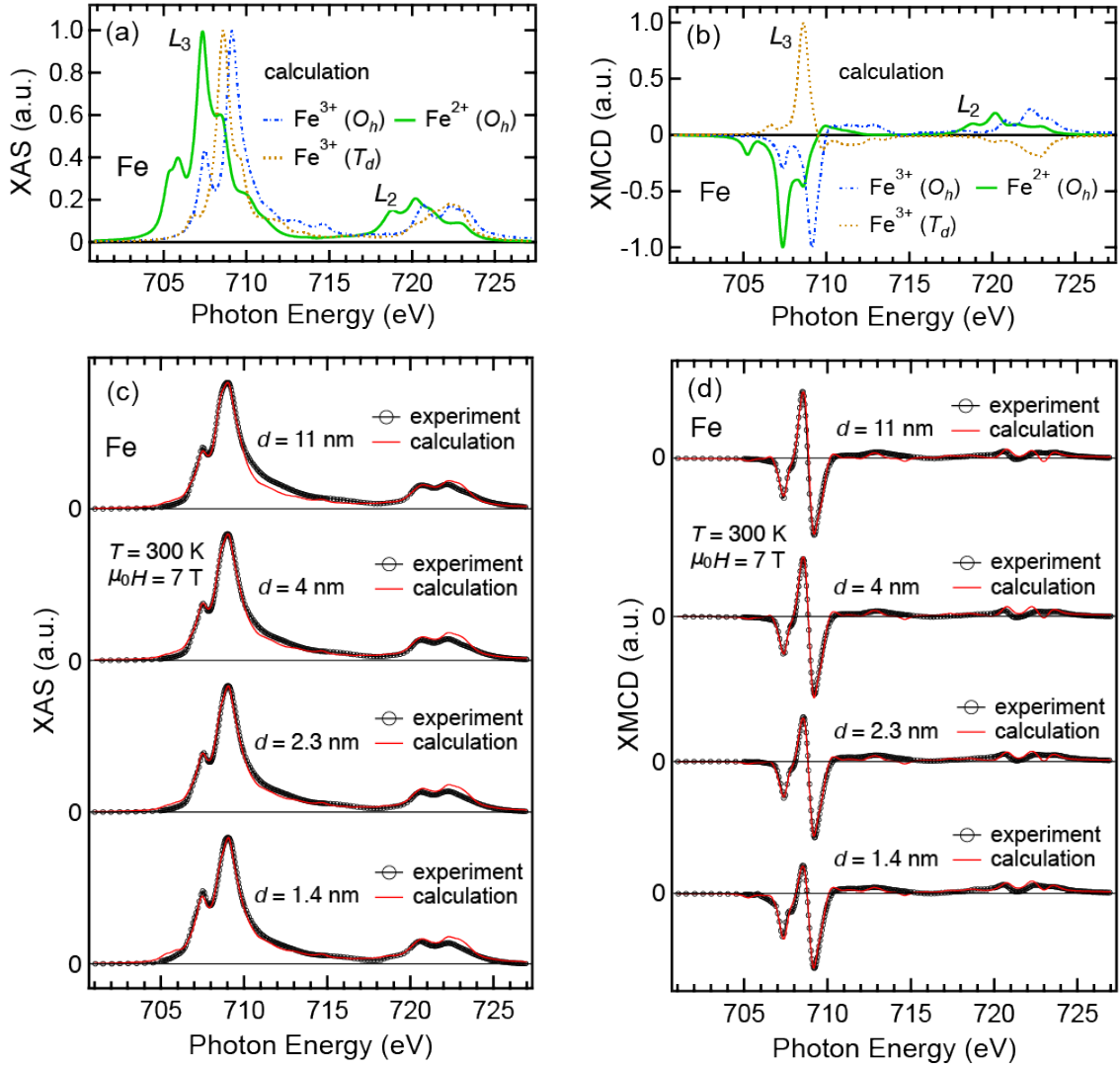

FIG. 7. (a)(b) Calculated Fe $L_{2,3}$-edge XAS [(a)] and XMCD [(b)] spectra, where the dot-dashed, dotted, and solid curves represent the spectra for $Fe^{3+}$ ($O_h$), $Fe^{3+}$ ($T_d$), and $Fe^{2+}$ ($O_h$), respectively. (c)(d) Experimental Fe $L_{2,3}$-edge XAS [(c)] and XMCD [(d)] spectra for the CoFe$_2$O$_4$ films with $d$ = 1.4, 2.3, 4, and 11 nm measured at 300 K with a magnetic field $\mu_0 H$ = 7 T applied perpendicular to the film surface. In the figure, the circles are experimental data and the red curves are the weighted sum of the calculated spectra shown in panels (a) and (b). Each spectrum has been arbitrarily scaled for easy comparison.

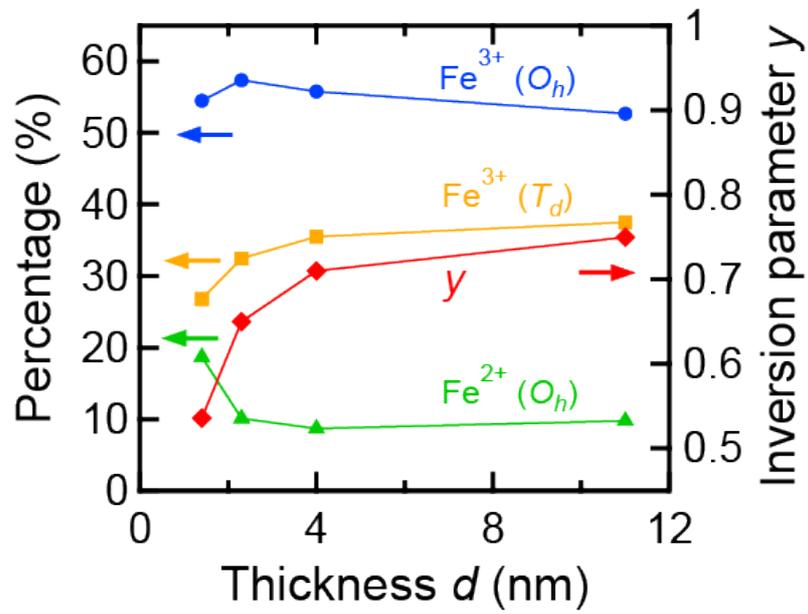

FIG. 8. Thickness $d$ dependence of site occupancies, where the circles, squares, and triangles represent $Fe^{3+}$ ($O_h$), $Fe^{3+}$ ($T_d$), and $Fe^{2+}$ ($O_h$), respectively. Inversion parameter $y$ is also shown as rhombuses.

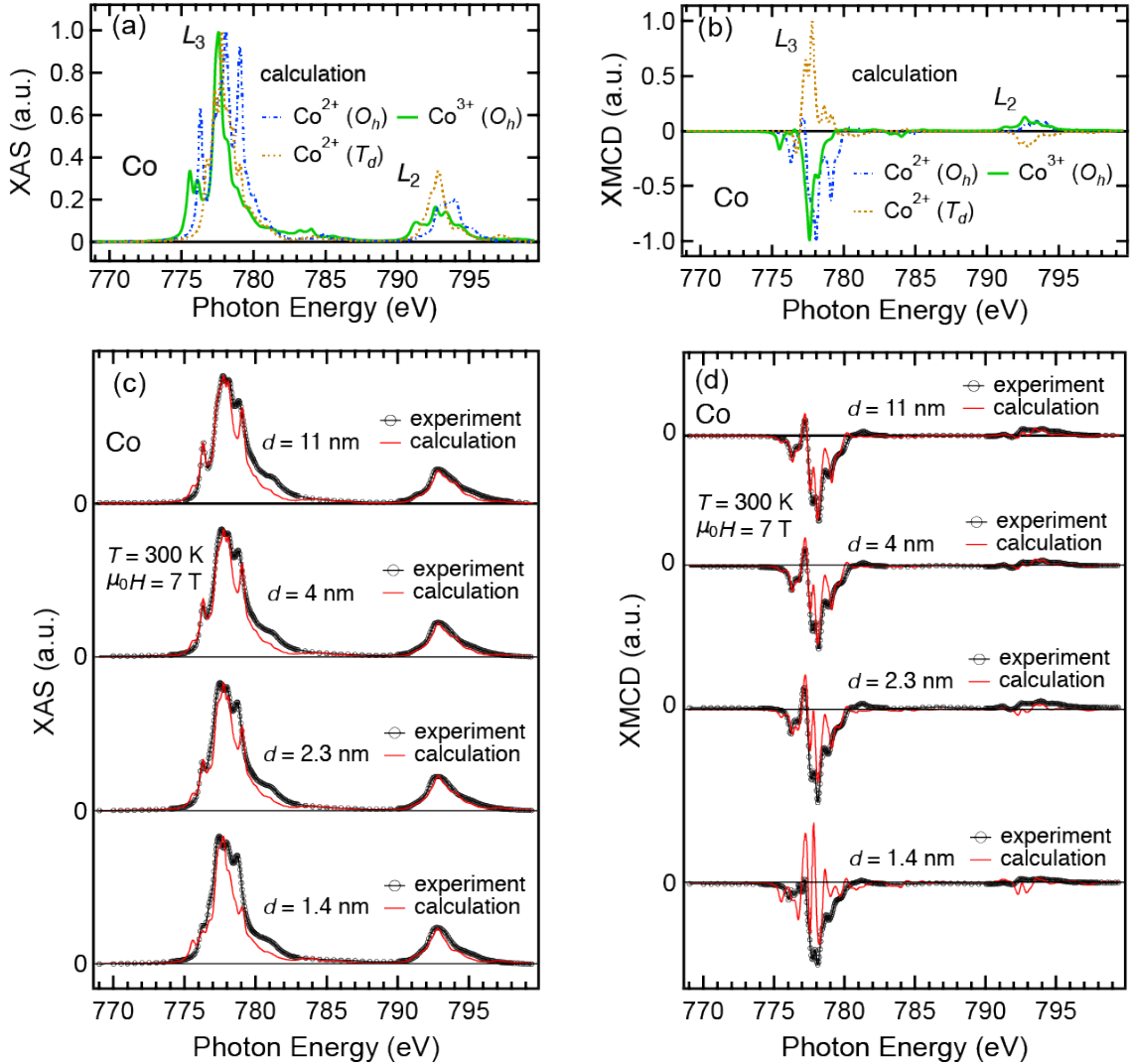

FIG. 9. (a)(b) Calculated Co $L_{2,3}$-edge XAS [(a)] and XMCD [(b)] spectra, where the dot-dashed, dotted, and solid curves represent $Co^{2+}$ ($O_h$), $Co^{2+}$ ($T_d$), and $Co^{3+}$ ($O_h$), respectively. (c)(d) Experimental Co $L_{2,3}$-edge XAS [(c)] and XMCD [(d)] spectra for the $CoFe_2O_4$ films with $d$ = 1.4, 2.3, 4, and 11 nm measured at 300 K with a magnetic field $\mu_0 H$ = 7 T applied perpendicular to the film surface. In the figures, the circles are experimental data and the red curves are the weighted sum of the calculated spectra shown in panels (a) and (b). Here, the site occupancies for the Co cations are estimated from that for the Fe cations shown in Fig. 8 under the conditions that the charge neutrality is maintained and that the number ratio of the $O_h$ sites to the $T_d$ sites is 2. Each spectrum has been arbitrarily scaled for easy comparison.

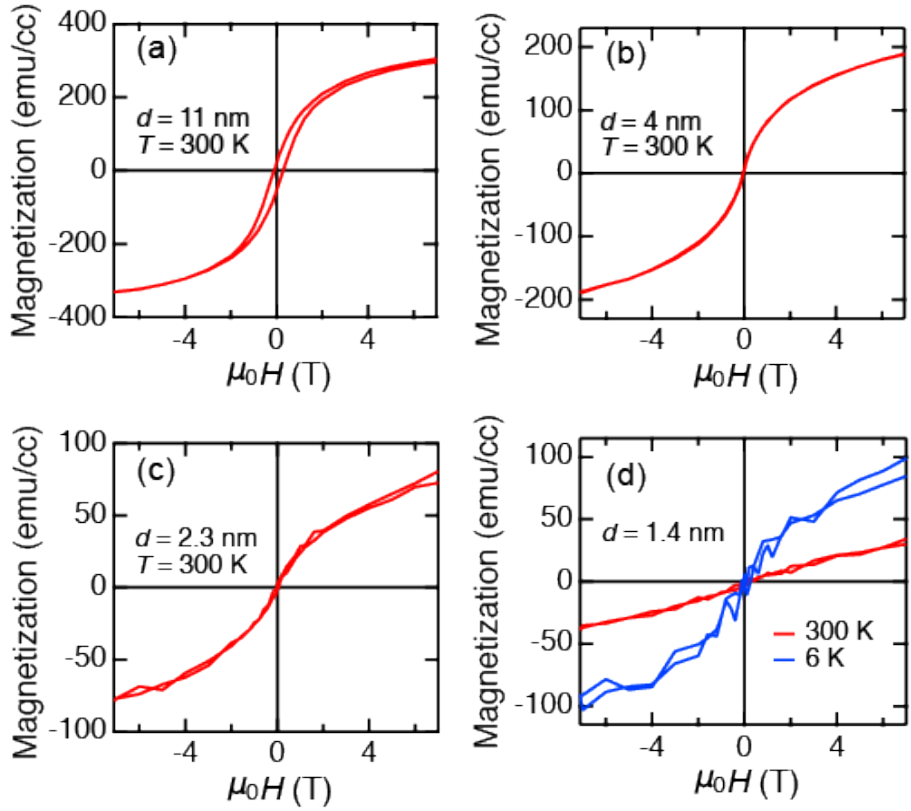

FIG. 10. (a)-(d) XMCD – H curves measured at the Fe $L_3$-edge for the CoFe$_2$O$_4$ films with $d$ = 11 [(a)], 4 [(b)], 2.3 [(c)], and 1.4 [(d)] nm, in which red curves and blue curves represent the signals measured at 300 and 6 K, respectively. The vertical axis of the XMCD intensity has been scaled so that it represents the sum of the magnetizations of the Fe and Co cations estimated from the fits in Figs. 6(c), 6(d), 8(c) and 8(d).

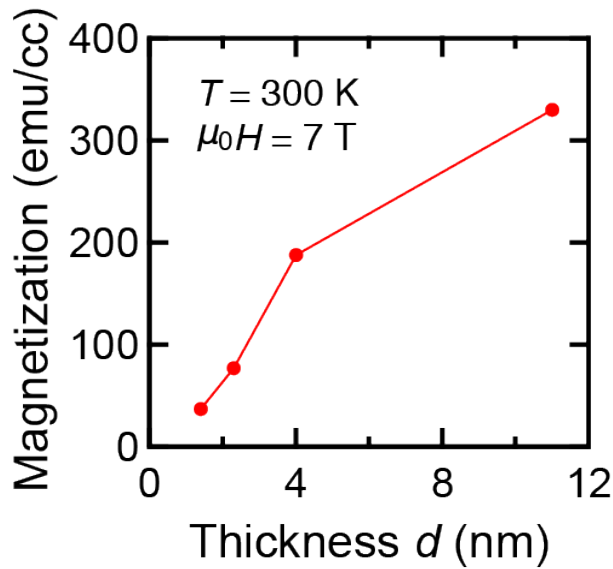

FIG. 11. Thickness ($d$) dependence of the sum of the magnetizations of the Fe and Co cations derived from the values of red curves at 7 T in Fig. 9.

|  | Δ | 10Dq | $pd\sigma$ | $U_{dd}$ |
|---|---|---|---|---|
| $Fe^{3+}$ ($O_h$) | 0.4 | 0.9 | 1.2 | 6.0 |
| $Fe^{3+}$ ($T_d$) | 4.0 | -0.5 | 2.0 | 6.0 |
| $Fe^{2+}$ ($O_h$) | 6.5 | 0.9 | 1.6 | 6.0 |
| $Co^{2+}$ ($O_h$) | 5.6 | 0.5 | 1.3 | 6.5 |
| $Co^{2+}$ ($T_d$) | 6 | -0.3 | 1.4 | 6.0 |
| $Co^{3+}$ ($O_h$) | 0 | 0.5 | 1.3 | 6.0 |

Table 1. Parameter values in units of eV used in the calculation based on the CI cluster model. For the Fe cations, $U_{dd}$ was adopted from [28].

|  | $Fe^{3+}$ ($O_h$) | $Fe^{3+}$ ($T_d$) | $Fe^{2+}$ ($O_h$) | $Co^{2+}$ ($O_h$) | $Co^{2+}$ ($T_d$) | $Co^{3+}$ ($O_h$) |
|---|---|---|---|---|---|---|
| $m_{spin}$ ($\mu_B$/atom) | 4.406 | 4.557 | 3.620 | 2.54 | 2.87 | 3.20 |
| $m_{orb}$ ($\mu_B$/atom) | 0.015 | 0.004 | 0.556 | 1.07 | 0.47 | 0.70 |

Table 2. $m_{spin}$ and $m_{orb}$ ($\mu_B$/atom) calculated within the CI cluster model for the $Fe^{3+}$ ($O_h$), $Fe^{3+}$ ($T_d$), $Fe^{2+}$ ($O_h$), $Co^{2+}$ ($O_h$), $Co^{2+}$ ($T_d$), and $Co^{3+}$ ($O_h$) cations.

# Supplemental Material
# Electronic structure and magnetic properties of magnetically dead layers in epitaxial CoFe$_2$O$_4$/Al$_2$O$_3$/Si(111) films studied by X-ray magnetic circular dichroism

Yuki K. Wakabayashi,[1] Yosuke Nonaka,[2] Yukiharu Takeda,[3] Shoya Sakamoto,[2] Keisuke Ikeda,[2] Zhendong Chi,[2] Goro Shibata,[2] Arata Tanaka,[4] Yuji Saitoh,[3] Hiroshi Yamagami,[3,5] Masaaki Tanaka,[1,6] Atsushi Fujimori,[2] and Ryosho Nakane[1,7]

[1]*Department of Electrical Engineering and Information Systems,*
*The University of Tokyo, 7-3-1 Hongo, Bunkyo-ku, Tokyo 113-8656, Japan*
[2]*Department of Physics, The University of Tokyo, Bunkyo-ku, Tokyo 113-0033, Japan*
[3]*Materials Sciences Research Center, Japan Energy Atomic Agency, Sayo, Hyogo 679-5148, Japan*
[4]*Department of Quantum Matters, ADSM, Hiroshima University, Higashi-Hiroshima 739-8530, Japan*
[5]*Department of Physics, Kyoto Sangyo University, Motoyama, Kamigamo, Kita-Ku, Kyoto 603-8555, Japan*
[6]*Center for Spintronics Research Network, Graduate School of Engineering,*
*The University of Tokyo, 7-3-1 Hongo, Bunkyo-ku, Tokyo 113-8656, Japan*
[7]*Institute for Innovation in International Engineering Education,*
*The University of Tokyo, 7-3-1 Hongo, Bunkyo-ku, Tokyo 113-8656, Japan.*

## I. Low-energy electron diffraction (LEED) and X-ray diffraction (XRD) measurements

The 6-fold symmetry of a 11-nm-thick CoFe$_2$O$_4$ layer with the 2×2 reconstruction was also confirmed by a low-energy electron diffraction (LEED) pattern, as shown in Fig. S1.

To characterize the crystalline properties, we carried out X-ray diffraction (XRD) measurements. To estimate the lattice constant $a_n$ along the growth direction normal to the film plane, the $\theta$-$2\theta$ method was used. To analyze the epitaxial relationship with respect to the Si substrate, the X-ray source and detector were fixed at the position for detecting specific planes such as (311), while the substrate plane was rotated. Here, the in-plane angle with respect to the origin is defined as $\phi$, and a XRD pattern measured by the latter method is referred to as the $\phi$-scan XRD pattern. In $\theta$-$2\theta$ XRD, the peaks assigned to γ-Al$_2$O$_3$(444) and CoFe$_2$O$_4$(444) were observed, indicating that the growth direction of both the γ-Al$_2$O$_3$ and CoFe$_2$O$_4$ layers was [111] on the Si(111) substrate. From these peaks, $a_n$ of the γ-Al$_2$O$_3$ layer was estimated to be larger by 6% than that of bulk materials. Thus, the γ-Al$_2$O$_3$ layer was tensile strained. On the other hand, $a_n$ of the CoFe$_2$O$_4$ layer was the same as that of bulk CoFe$_2$O$_4$. Figure S2 shows $\phi$-scan XRD patterns for Si(311), γ-Al$_2$O$_3$(440), and CoFe$_2$O$_4$ (311). As expected from the RHEED and LEED patterns, the γ-Al$_2$O$_3$ and CoFe$_2$O$_4$ layers have the 6-fold symmetry, in which one domain is completely aligned with the Si substrate with the epitaxial relationship γ-Al$_2$O$_3$[11$\bar{2}$](111) // CoFe$_2$O$_4$[11$\bar{2}$](111) // Si[11$\bar{2}$](111), whereas another domain is rotated by 60° in the (111) plane. This double domain structure is basically the same as that in Ref. [S1].

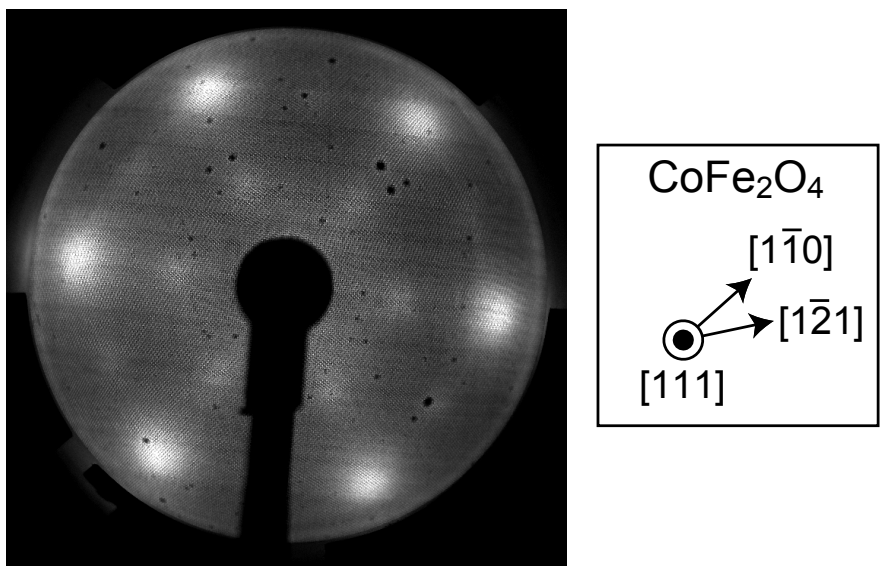

FIG. S1. Low-energy electron diffraction (LEED) pattern of an epitaxial 11-nm-thick $CoFe_2O_4$ layer grown on an epitaxial $\gamma$-$Al_2O_3$(111)/Si(111) structure, which was taken with the electron energy of 80 eV. The crystalline directions of the Si substrate are also shown.

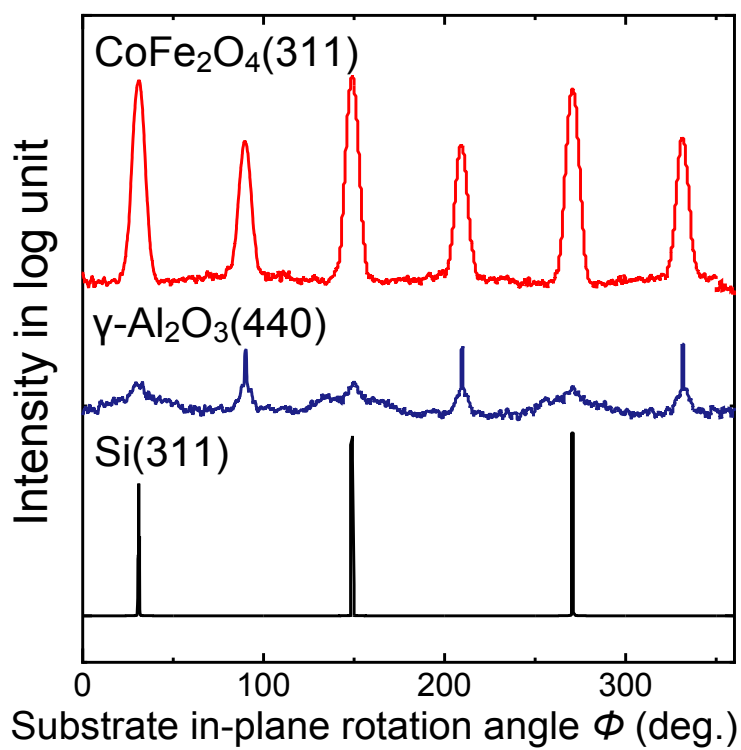

FIG. S2. $\phi$ scan X-ray diffraction pattern of an epitaxial $CoFe_2O_4$/$\gamma$-$Al_2O_3$(111)/Si(111) structure, in which the black, blue, and red curves are the patterns taken from Si(311), $\gamma$-$Al_2O_3$(440), and $CoFe_2O_4$(311) planes, respectively. The magnitude of each data is adjusted for easy comparison.

## II. X-ray absorption spectroscopy (XAS) and X-ray magnetic circular dichroism (XMCD) spectra for the CoFe$_2$O$_4$ films with the thickness $d$ = 1.5, 2.5 and 4.3 nm.

Figure S3 shows the Fe and Co $L_{2,3}$-edge XAS [$\mu^-$, $\mu^+$, and $(\mu^+ + \mu^-)/2$] and XMCD $(\mu^+ - \mu^-)$ spectra for the CoFe$_2$O$_4$ film with $d$ = 1.4 nm at 300 K with magnetic fields $\mu_0 H$ = 1 and 7 T applied perpendicular to the film surface. The spectra with $\mu_0 H$ = 1 T were measured after the application of $\mu_0 H$ = 7 T. The insets in Figs. S3(c) and (d) show the magnified plots of the XMCD spectra at the $L_3$-edges normalized at 709.2 and 778.2 eV, respectively. Figures S4 and S5 show the same date measured for the CoFe$_2$O$_4$ films with $d$ = 2.3 and 4 nm, respectively. In all the samples, for both Fe and Co $L_3$-edges, the normalized XMCD spectra with various $H$ are identical with each other, which confirms that all the CoFe$_2$O$_4$ films with various thicknesses are made of a magnetically single phase.

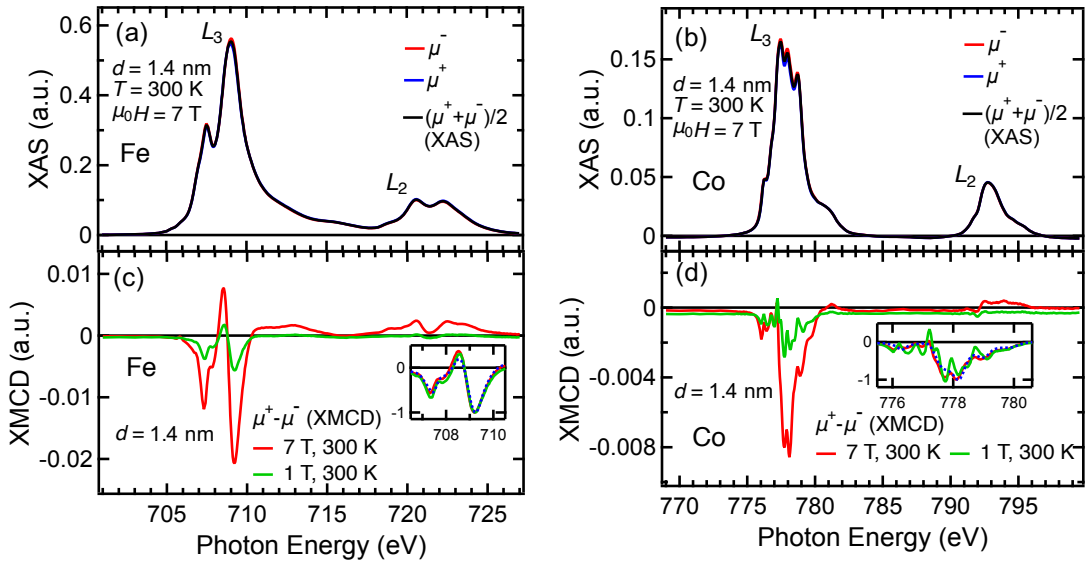

FIG. S3. (a) Fe and (b) Co $L_{2,3}$-edge XAS [$(\mu^+ + \mu^-)/2$] spectra for the CoFe$_2$O$_4$ film with $d$ = 1.4 nm which were measured at 300 K with a magnetic field $\mu_0 H$ = 7 T applied perpendicular to the film surface. (c) Fe and (d) Co $L_{2,3}$-edge XMCD ($= \mu^+ - \mu^-$) spectra for the CoFe$_2$O$_4$ film with $d$ = 1.4 nm which were measured at 6 and 300 K with magnetic fields $\mu_0 H$ = 1 and 7 T applied perpendicular to the film surface. The insets in (c) and (d) show a magnified plot of the spectra at the $L_3$-edges normalized at 709.2 and 778.2 eV, respectively.

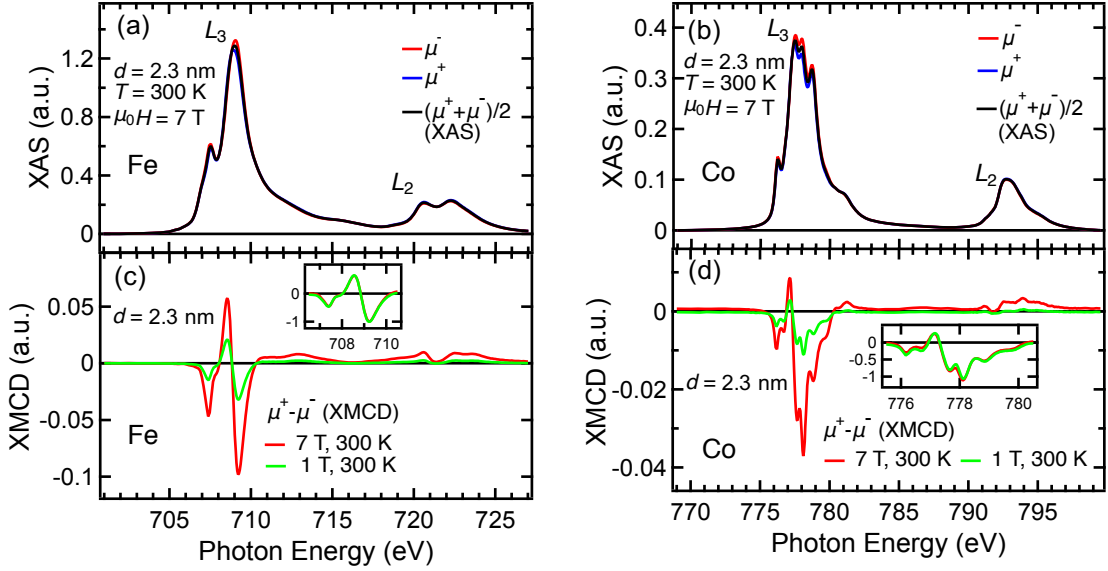

FIG. S4. (a) Fe and (b) Co $L_{2,3}$-edge XAS $[(\mu^+ + \mu^-)/2]$ spectra for the CoFe$_2$O$_4$ film with $d = 2.3$ nm which were measured at 300 K with a magnetic field $\mu_0 H = 7$ T applied perpendicular to the film surface. (c) Fe and (d) Co $L_{2,3}$-edge XMCD ($= \mu^+ - \mu^-$) spectra for the CoFe$_2$O$_4$ film with $d = 2.3$ nm which were measured at 6 and 300 K with magnetic fields $\mu_0 H = 1$ and 7 T applied perpendicular to the film surface. The insets in (c) and (d) show a magnified plot of the spectra at the $L_3$-edges normalized at 709.2 and 778.2 eV, respectively.

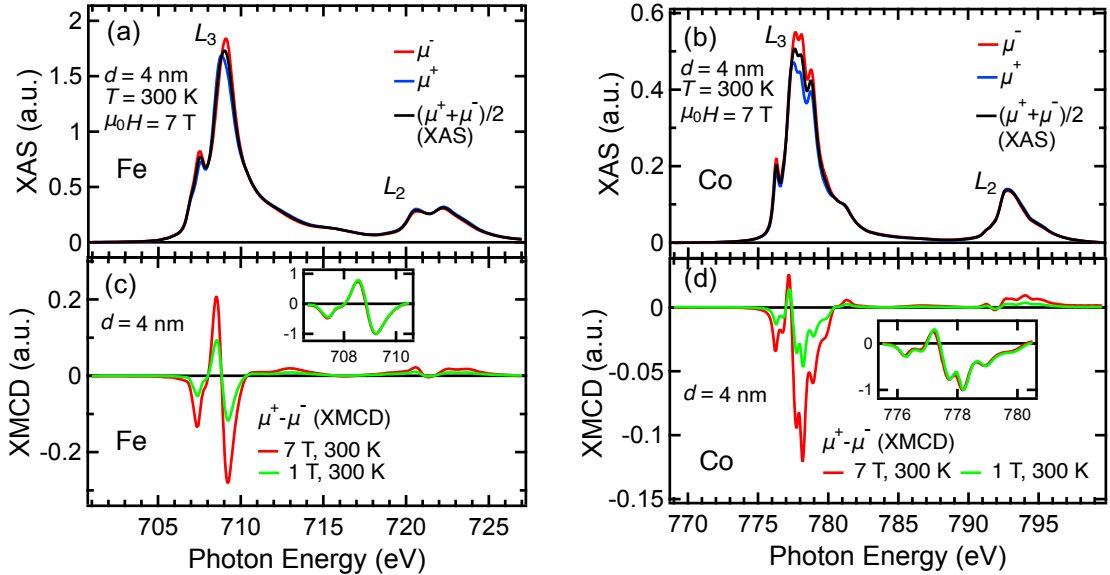

FIG. S5. (a) Fe and (b) Co $L_{2,3}$-edge XAS $[(\mu^+ + \mu^-)/2]$ spectra for the CoFe$_2$O$_4$ film with $d = 4$ nm which were measured at 300 K with a magnetic field $\mu_0 H = 7$ T applied perpendicular to the film surface. (c) Fe and (d) Co $L_{2,3}$-edge XMCD ($= \mu^+ - \mu^-$) spectra for the CoFe$_2$O$_4$ film with $d = 4$ nm which were measured at 6 and 300 K with magnetic fields $\mu_0 H = 1$ and 7 T applied perpendicular to the film surface. The insets in (c) and (d) show a magnified plot of the spectra at the $L_3$-edges normalized at 709.2 and 778.2 eV, respectively.

## III. XMCD - $H$ curves measured at the Fe and Co $L_3$-edges

Figures S6(a)-(d) show the XMCD - $H$ curves at the Fe and Co $L_3$-edges for the CoFe$_2$O$_4$ films with $d$ = 11, 4, 2.3, and 1.4 nm, respectively, in which the blue curves and the red curves are the results measured for the Fe and Co $L_3$-edges, respectively. Here, the XMCD curves are normalized at $\mu_0 H$ = 7 T. In all the samples, the normalized XMCD - $H$ curves for both Fe and Co $L_3$-edges are identical, which confirms again that the XMCD signals come from the single magnetic phase.

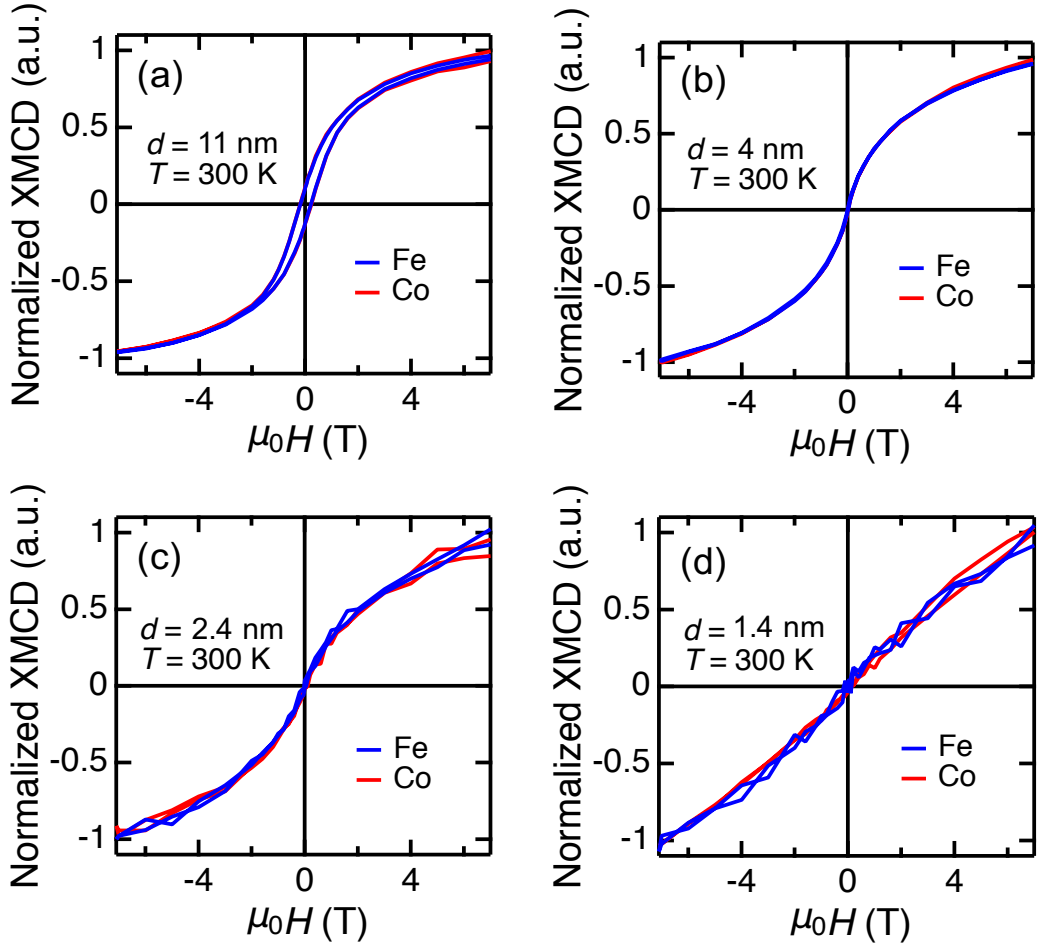

FIG. S6. (a)-(d) XMCD - $H$ curves measured at the Fe and Co $L_3$-edges for the CoFe$_2$O$_4$ films with $d$ = (a) 11, (b) 4, (c) 2.3, and (d) 1.4 nm at 300 K in which the blue curves and red curves represent the signals measured for the Fe and Co $L_3$-edges, respectively. The XMCD curves are normalized at $\mu_0 H$ = 7 T.

**Reference**

[S1] R. Bachelet, P. de Coux, B. Warot-Fonrose, V. Skumryev, G. Niu, B. Vilquin, G. Saint-Girons, and F. Sánchez, CrystEngComm **16**, 10741 (2014).